\documentclass[12pt, reqno]{amsart}
\usepackage{graphicx}
\usepackage[hyphens]{url}

\theoremstyle{plain}

\theoremstyle{remark}

\title{Proof Simplification and Automated Theorem Proving}
\author{Michael Kinyon}
\thanks{${}^*$ Partially supported by Simons Foundation Collaboration Grant 359872 and
by FCT project CEMAT-CI\^{E}NCIAS UID/Multi/04621/2013}
\address{Department of Mathematics, University of Denver, Denver, CO 80208, USA}
\email{mkinyon@du.edu}

\begin{document}

\begin{abstract}
The proofs first generated by automated theorem provers are far from optimal by
any measure of simplicity. In this paper I describe a technique for simplifying automated proofs.
Hopefully this discussion will stimulate interest in the larger, still open, question of what reasonable measures
of proof simplicity might be.
\end{abstract}

\maketitle

\section{A Personal Introduction}

Although I am neither a professional historian nor philosopher, in the summer of 2000, I attended the
Annual Meeting of the Canadian Society for the History and Philosophy of Mathematics at McMaster University
in Ontario. Considering the subject of this paper, it is a remarkable coincidence that the first plenary talk
of that conference was R\"{u}diger Thiele's announcement of finding the lost Hilbert's 24th problem.
The following image is taken directly from the book of abstracts for that meeting \cite{CSHPM}

\begin{center}
  {\includegraphics[width=11.5cm]{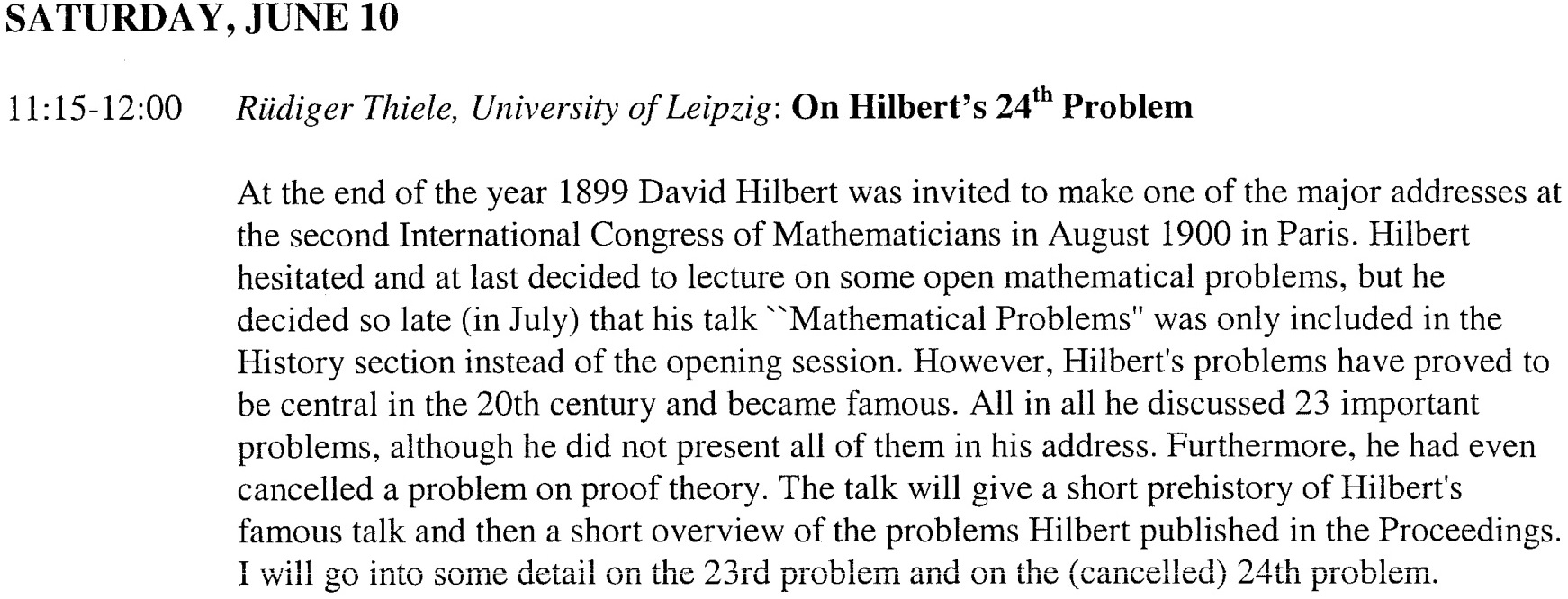}}
\end{center}

For reasons I can no longer recall, I also volunteered to be the Editor of that conference's Proceedings,
and so I had the distinct pleasure of editing Prof. Thiele's contribution \cite{Thiele1}.

At the same time, I was in the process of changing my mathematical research speciality, turning from
differential equations to nonassociative algebra. One month after attending the aforementioned conference,
I and two coauthors submitted a paper \cite{KKP} in which the main theorem was obtained via the assistance of
OTTER, an automated theorem prover \cite{OTTER}. That paper changed my career, and since then nearly all of
my research has used automated theorem proving to obtain results in mathematics.

There are many automated theorem provers (ATPs) for an interested mathematician to choose from. The most popular ones
include \textsc{Prover9} \cite{McCune}, \textsc{E} \cite{Schulz}, \textsc{Waldmeister} \cite{Hillenbrand}
and \textsc{Vampire} \cite{Voronkov}. Most of my own experience has been with \textsc{Prover9} and in fact,
only it and \textsc{E} have the necessary infrastructure to implement the proof simplification techniques
I will describe later. Another advantage of \textsc{Prover9} for mathematicians is that its learning curve
is less steep and unlike the others, its input language is easily readable by humans.

Automated theorem provers search for proofs by generating and processing inferences much faster than humans can.
If a proof of a goal is found, the prover reports a successful search and in most cases, presents the proof.
An ATP proof consists of a sequence of clauses -- where each clause is a conjunction of atomic formulas
($n$-ary predicates applied to $n$ terms) or their negations -- and the justification for each clause,
that is, the inferences used to derive the clause from its parents.

An important point here is that an ATP will report the \emph{first} proof it finds. This is rarely an optimal
proof by any measure of optimality. In particular, the first proof found is never (in my experience) a \emph{simple}
proof.

What exactly \emph{is} a ``simple'' proof and what makes one proof simpler than another? This is a deep
question which I am not going to attempt to answer here. The most commonly used measure of simplicity
\cite{ThieleWos,Veroff1,WosThiele} is \emph{proof length}, which is the number of inference
steps. Thus efforts to find simpler proofs of theorems usually focus on finding \emph{shorter} proofs.

I will follow the same path here: in this paper, simplification will mean looking for short proofs. By way of
offering a caveat, let me mention some other measures (already discussed by Thiele and Wos \cite{ThieleWos,WosThiele})
which a fully developed theory of proof simplicity for ATP proofs should take into account.

A proof can be visualized in various ways. The easiest is just to think of it as a sequence of clauses, and
this is essentially how the ``simple equals short'' point of view sees them. However, they can also be visualized
as directed graphs, where each vertex is a clause and there is an arrow from one clause to another if the first
is an immediate parent of the second. In reality, a proof graph can be quite complex, with many paths coming
out of the axiom clauses. Where those paths join up later in the proof are usually important clauses.
Thus one measure of proof simplicity which deserves to be studied in more detail is the complexity of the proof
graph, because generally speaking, a proof with fewer paths emanating from the axioms will be intuitively simpler
than a proof with many such paths. It is worth noting, however, that this measure of simplicity is rather well
correlated with proof length, because when the clauses in a proof are arranged in a sequence, fewer paths generally
means fewer inference steps.

Other measures of proof simplicity are related to the \emph{weights} of clauses. The weight of a clause is
the number of symbols which occur in it, not counting parentheses. For instance, the associative law
$(x\ \ast\ y)\ \ast\ z\ =\ x\ \ast\ (y\ \ast\ z)$ has weight $11$ (note that the equality symbol is counted).
Generally speaking, a proof containing many clauses of high weight is less simple than a proof with fewer
such clauses. So measures of proof simplicity might include the \emph{maximum} clause weight occuring in
a proof or the \emph{average} clause weight of a proof, and so on.

Automated theorem provers are resolution-based and as such, always prove goals by contradiction. (Resolution is an inference rule 
which can be seen as a generalization of modus ponens. It essentially takes two clauses which contain common literals, positive in one clause, negative in the other, and produces a new clause consisting of the remaining literals.) So for instance, if the universally quantified commutative law $x\ \ast\ y\ =\ y\ \ast\ x$
is the goal, \textsc{Prover9} will first convert this goal into a denial $c1\ \ast\ c2\ !=\ c2\ \ast\ c1$, which
can be read as ``there exist $c1, c2$ such that $c1\ast c2$ is not equal to $c2\ast c1$.''
An ATP might derive a goal directly or it might partially reason backwards from the denial
until some negative clause contradicts a positive clause. There is a general sense among
mathematicians that forward proofs, going from the hypotheses to the conclusions, are
``better'' than backward proofs by contradiction, and especially better than proofs that mix the two.
By extension, an ATP proof with a lot of backward steps might not be considered as ``good'' as a proof
with very few, if any, such steps. Should a strictly forward proof be considered simpler?

There is a delicate trade-off between all these measures. For example, suppose that one proof is precisely
one clause shorter than another, but the slightly shorter proof has a significantly higher maximum clause weight.
Is the shorter proof really ``simpler'' than the other one? An answer to this question is obviously context-dependent.
In the spirit of Hilbert's own research programme, and what may have motivated his thinking for the cancelled 24th problem,
finding simpler proofs seems to be something like a \emph{variational problem} where the constraints are based on
some balance between the various possible measures of proof simplicity. In short, as Hilbert realized, the issue of proof
simplicity is not so simple.

The simplification technique I am going to describe in the rest of this paper still focuses on proof length.
However, in my experience, the other measures I have mentioned tend to simplify as well (though this does not
always happen). In order to make sense of the technique, I must first describe in the next section the
given clause algorithm that underlies essentially most ATP systems. Following that, I will give a brief outline
of Veroff's method of proof sketches \cite{Veroff2}. In the penultimate section I will discuss the proof
simplification technique, and finally I will sum up with some concluding remarks.

\section{The Given Clause Algorithm}

In order to make sense of how I use \textsc{Prover9} for proof simplification, I must first
discuss its main loop, the basics of which are the same for most ATPs. So let us assume
all preprocessing has been completed, the user's input has been transformed into clauses, and the
system is now ready to make inferences and search for a proof.
There are three key objects in the inference process:
\begin{itemize}
  \item the \emph{given clause}
  \item the \emph{SOS} (set of support)
  \item the \emph{usable list}
\end{itemize}
The \emph{given clause} is the important one here because it controls what
inferences can be made. It will make more sense after I describe the other two objects.

The \emph{SOS} is just the list of clauses that are waiting to be selected as given clauses.
SOS clauses are not available for making primary inferences, but in some systems can be
used to rewrite other clauses. The size of the SOS is the fundamental measure of the size
of the search space: a large SOS is a sign of a complex search that may take a long time
to find a proof.

The \emph{usable list} is the list of clauses that are available for making
primary inferences with the given clause. At each iteration of the main loop, a given clause
is selected from the SOS, moved to the usable list, and then inferences are made using the
given clause and other clauses already in the usable list. More precisely, the algorithm
can be summarized as a while loop.

\bigskip

\noindent\texttt{while the SOS is not empty:
\begin{enumerate}
\item Select a given clause from SOS and move it to the
    usable list;
\item Infer new clauses using the inference rules in effect;
       each new clause must have the given clause as one of its
       parents and members of the usable list as its other
       parents;
\item process each new clause;
\item append new clauses that pass the retention tests to the SOS.
\end{enumerate}
end of while loop.}

\bigskip

There are many retention tests. An obvious one is that
if a newly generated clause is subsumed by a clause already on the SOS, then
there is no reason to keep the new one.
(One clause subsumes another if the variables of the first can be
instantiated in such a way that it becomes a subclause of the second.) Others might be based on practical
concerns; for instance, clauses of exceptionally high weight might be discarded
to keep the size of the SOS manageable.

The most important aspect of this loop is the selection of the given clause
from the SOS. The choice of selection strategy very much determines how the
search will go, and greatly affects the chances of the system finding a proof.

An example is a \emph{breadth first} search. This just selects the oldest available
clause in the SOS to be the given clause, where ``oldest'' means that it was
kept in the SOS before all other clauses. Another example is a \emph{lightest first} search,
which selects the clause of least weight among all clauses available in the SOS.
If there are multiple clauses with the same weight, then the oldest among that subset
is selected.

Most hueristic strategies are a mix of the above: select an old clause, then a few light clauses,
then an old clause, and so on. This is sometimes called a \emph{ratio strategy}, referring to
the ratio of old clauses to light clauses.

Prover9's default selection scheme works in a cycle of 9 steps:
\begin{itemize}
\item Select the oldest available clause
\item Select the four lightest available ``false'' clauses
\item Select the four lightest available ``true'' clauses
\end{itemize}
For this exposition, one may think of ``false'' as coming from reasoning backward from the
denial and ``true'' as coming from strictly forward inferences. (This is not quite the
precise meaning for \textsc{Prover9}, but the details are not important for what follows.)
The values ``$1$ old, $4$ true, $4$ false'' were chosen by developer Bill McCune experimentally.
In many equational problems, that is, \textsc{Prover9} jobs where equality is the only
predicate, there are very few false clauses available, so the scheme effectively becomes
``$1$ old, $4$ true''. These default numbers can be easily changed by the user. Other
ATPs such as \textsc{E} also allow customization of the given clause selection rules.

\section{Proof Sketches and Hints}

In this section I will describe a particularly powerful method for selecting given
clauses which was developed by Veroff \cite{Veroff2}. This is called the method of
\emph{proof sketches}. It was originally implemented in OTTER (as \emph{hint lists}),
and is implemented in two current systems, \textsc{Prover9} (again as hint lists) and
\textsc{E} (as watchlists). The general principles apply to any implementation.

The basic idea behind proof sketches is the following: suppose the desired theorem
is easier to prove in a special case than it is in a more general case. It might
nevertheless hold that the proofs of the theorem in both the special and general cases
have certain features in common. The idea is to try to ``guide'' the ATP system in
the direction of the special case's proof, called a \emph{proof sketch}, as it searches
for a proof in the general  case.

To make this a little more concrete, suppose the goal is a theorem about distributive lattices,
and suppose further that the theorem is easy to prove in Boolean algebras. One might get the ATP
to prove the theorem for Boolean algebras, extract the clauses from the proof, and then (informally
speaking) have the ATP refer to that proof as it selects given clauses in the search for the
more general case of distributive lattices. That is, any time a clause is available in the SOS
which ``looks like'' a clause in the already found proof, that SOS clause should have high priority
to be selected as a given clause.

Again, think of a proof as a sequence of clauses:
\[
C_1 , C_2, \ldots, C_i \ldots, C_j \ldots, C_n\,.
\]
Suppose that $C_i$ is an extra assumption in the target theory, not an assumption that one desires
to have in the following. In the aforementioned example, $C_i$ might be a clause axiomatizing
Boolean algebras among distributive lattices. Suppose further that clause $C_j$ has clause $C_i$
in its derivation history, meaning that $C_i$ is an ancestor of $C_j$.

Now one of two things can happen. First, $C_j$ itself might be in the target theory, for instance,
in our example, $C_j$ might hold in all distributive lattices even though in this particular proof, its
derivation depended upon the extra assumption $C_i$. If this occurs, then to complete (part of)
the proof of the general case, we need to find a new proof of $C_j$ which does not depend on
$C_i$. Alternatively, $C_j$ itself might not be in the target theory, so in our example, $C_j$ might
be true in Boolean algebras but not distributive lattices. In this case, we need to bridge the
gap between the assumptions without $C_i$ and some clause occurring after $C_j$ which holds in the
target theory. In either case, we have hopefully reduced a large problem to a smaller one, at least
in principle.

As mentioned above, the proof sketches strategy is implemented in two current ATP systems,
\textsc{Prover9}'s hint list and \textsc{E}'s watchlist. The basic idea behind both implementations
is essentially the same: when a new clause is generated, the ATP checks if the new clause
\emph{matches} a \emph{hint} clause on the list. Here a clause is said to \emph{match} a hint
if it subsumes it. When this occurs, the new clause is put immediately in the SOS and is marked
as a \emph{hint matcher}.

Now hints are assumed to be important clauses, notable milestones toward a proof, because they
already occurred in another proof of a stronger theory. Thus the rest of the implementation
of proof sketches is to give hint matchers higher priority in given clause selection by
using the following rule: \emph{If there is a hint matcher in the SOS, select it as a given clause
as soon as possible}. If there are multiple hint matchers in the SOS (and this usually happens in
practice), then the rule has to be refined to allow the ATP to decide which hint matcher to
select first. Both \textsc{Prover9} and \textsc{E} default to selecting hint matchers first by
lowest weight and then by age. (In both case, the user can change this if desired.)

Many of the successful applications of ATP systems to real mathematical problems have used
proof sketches to find proofs. It has turned out to be a powerful technique in that regard.

\section{Proof Simplification}

Now let me turn the main theme: how proof sketches can be used in proof simplification.
I will focus specifically on \textsc{Prover9} this time.

Suppose that \textsc{Prover9} finds a proof, we extract a proof sketch from the proof into a hints
list, and then run \textsc{Prover9} again \emph{without changing the input} or any other settings.
Intuitively, we might expect \textsc{Prover9} to find the exact same proof as before.
In practice, that is not what happens, and a moment's thought tells us why: the presence of the
new hints means that hint matchers in the SOS are being selected as given clauses in a different
order than before. The ``butterfly effect'' of this change to the selection of given clauses
has sent the search off in a different direction.

Let us call an ATP proof \emph{stable} if, when the proof's hints are used as a sketch, then
leaving the input and all other settings unchanged, \textsc{Prover9} finds (essentially) the
exact same proof. What years of experience have taught me is the following:

\begin{center}
\emph{The first proof an ATP finds is usually very complex and very unstable!}
\end{center}

To see what happens, let us look at a concrete example, which is essentially the main
theorem of \cite{KKP}. Here is what the assumptions look like in \textsc{Prover9} syntax.
\bigskip

\begin{verbatim}
% inverse property loop axioms
e * x = x. x * e = x.
x' * (x * y) = y.
(x * y) * y' = x.

% inner mappings
T(x,y) = y' * (x * y).
L(x,y,z) = (z * y)' * (z * (y * x)).
R(x,y,z) = ((x * y) * z) * (y * z)'.

% A-loop axioms
T(x,y) * T(z,y) = T(x * z,y).
L(x,y,z) * L(u,y,z) = L(x * u,y,z).
R(x,y,z) * R(u,y,z) = R(x * u,y,z).
\end{verbatim}

It is not especially important here what these assumptions mean, because that is not
what we are going to focus on. Here is the goal.

\begin{verbatim}
% Moufang identity
((x * y) * x) * z = x * (y * (x * z)).
\end{verbatim}

If we run \textsc{Prover9} on this input using (almost) default settings,
it finds a proof. Here are some of the basic statistics of the proof.

\begin{verbatim}
% Length of proof is 470.
% Level of proof is 43.
% Maximum clause weight is 87.
% Given clauses 935.
\end{verbatim}

The length of the proof is the number of primary inferences. In this case, a unit equality
problem, the only primary inference rule is paramodulation, which is essentially just equality substitution: when
one side of one equation is inserted into terms of another equation. There are also
secondary inferences which are rewrites. \textsc{Prover9} does not count these as part
of the proof length by default, and for our purposes, this distinction is not important.

The level of the proof is the length of the longest chain of inferences from the empty
clause (reached when the contradiction is found) up to an input clause. The maximum
clause weight has been discussed, but notice that it is a quite large number. The
proof was found after 935 clauses had been selected as given. The time it took to find
the proof is immaterial because that can vary considerably on different machines.

For this demonstration, we are going to focus mostly on the length. We will consider
one proof to be simpler than another if the former has shorter length than the latter.
However, we will not completely ignore the other parameters, as will be seen shortly.

We now extract the 470 steps of the proof into a hints list. (There is a utility so that
this does not need to be done by hand.) We run \textsc{Prover9} again, changing nothing
else in the input. Recall that now the search space is different: some clauses are going
to be marked as hint matchers and selected as given clauses earlier than they were before.
Here are the statistics from the second proof.

\begin{verbatim}
% Length of proof is 265.
% Level of proof is 42.
% Maximum clause weight is 49.
% Given clauses 169.
\end{verbatim}

Notice that \emph{every} parameter dropped considerably. This new proof is not only
simpler by the basic measure of length but also by the other measures as well. We can
see that the hint matchers dominated the search by the fact that the proof was found
after only 169 given clauses were generated. To focus on what I think is important here, I will no
longer report the given clause number; it continues to drop, but does not change
significantly after a few iterations of this procedure.

We repeat the process, extracting the 265 clauses from this proof. In this demonstration
we will discard the previous set of hints, but keeping it can also sometimes lead to
interesting results. Here are the statistics from the third proof.

\begin{verbatim}
% Length of proof is 236.
% Level of proof is 39.
% Maximum clause weight is 49.
\end{verbatim}

The drop in the parameters is not as dramatic as before, but we still have a simpler
proof. Notice that the maximum clause weight, which started at 87, did not change in this iteration. We could continue in
the same vein for a while, but what will happen eventually is that the proof lengths
will no longer drop, but might in fact rise a bit. Generally speaking, what tends
to happen is that proof lengths (and the other parameters) rise and fall. A stable
proof might be found at some point or it might not.

Instead, let me introduce another part of this general technique: squeezing the
maximum clause weight. We reduce the allowed maximum clause weight of the search to 48.
This means that any time \textsc{Prover9} generates a clause of weight greater than 48,
it will be deleted instead of being put on the SOS. (We also have to set a flag that
tells \textsc{Prover9} to apply this weight restriction to hint matchers; by default,
it does not do so.) So what we expect to happen here is that the clause of weight 49
which showed up in the last two proofs will be generated but deleted before it has
a chance to be selected as given. If all goes well, \textsc{Prover9} will find a
proof that bypasses that clause. Here is the result.

\begin{verbatim}
% Length of proof is 192.
% Level of proof is 33.
% Maximum clause weight is 41.
\end{verbatim}

Notice that we had a very significant drop in the proof length, and in fact, the
maximum clause weight dropped well below our cutoff of 48. \textsc{Prover9} was
forced to search among clauses of smaller weight and it paid off.

Let us try again, reducing the maximum clause weight to 40. As before, we keep using
each proof as a sketch for the next proof.

\begin{verbatim}
% Length of proof is 216.
% Level of proof is 35.
% Maximum clause weight is 39.
\end{verbatim}

We see that we have lost ground, the previous proof is simpler than this one.
Knowing that we can always fall back on the previous proof if need be, let us
nevertheless continue with the current one, squeezing the maximum clause weight
to 38.

\begin{verbatim}
% Length of proof is 205.
% Level of proof is 33.
% Maximum clause weight is 34.
\end{verbatim}

At this point, we can continue to squeeze the maximum clause weight, but \textsc{Prover9}
will always report 34. This is because two of the input clauses above have weight 34, and
\textsc{Prover9} never applies weight restrictions to input clauses. Let us squeeze down
to 33.

\begin{verbatim}
% Length of proof is 195.
% Level of proof is 29.
\end{verbatim}

Here we are getting very close to the simplest proof we found above, but now a question is
raised. The earlier simple proof is only 3 steps shorter than this one, but this one has
a significantly smaller maximum clause weight (in fact, the noninput clauses have
weight no greater than 33). Perhaps this proof should be considered to be simpler than the
earlier one?

Let us bypass this question by squeezing the maximum clause weight to 30.

\begin{verbatim}
% Length of proof is 183.
% Level of proof is 30.
\end{verbatim}

Note that the level went up slightly, but the length is down quite a bit, so it is reasonable
to say that this proof is simpler than any that came before.

We continue in this same vein, with proof lengths going up and down.
Squeezing the maximum clause weight below 24 turns out to cause \textsc{Prover9} not to find
any proof at all, instead the SOS becomes empty. At that point, we just leave the maximum
clause weight at 24 and continued iterating the basic procedure. Eventually we find a proof
of length 150, level 40. Although the level has gone up, the drop in the length suggests that
this proof is simpler than the previous one.

The maximum clause weight is only one \textsc{Prover9} parameter we can tweak. We can also
force \textsc{Prover9} to select hints by age instead of by weight, and this too shakes up
the search space. I did this once and then reverted to the default selection scheme.
After a couple of more iterations, \textsc{Prover9} found a proof of length 150, level 29.
This is the simplest proof I was able to find in this casual hunt for one.

There are many other things that can be done as well, for instance, force \textsc{Prover9}
to select some clauses which do not match hints before it runs out of hint matchers.
Another parameter that can be squeezed is the maximum number of variables. And there are
others as well. All of them are worth trying.

The whole process can be easily automated, of course. I carried out the above demonstration
``by hand'' to explain the basic technique, but there is no reason not to let the procedure
run by itself for a long time and then see what is the simplest proof it found in the whole
endeavor.

I should also mention one point I ignored above in my efforts to simplify the exposition.
As mentioned, \textsc{Prover9} does not count secondary inferences (rewrites) when it reports
proof length. It is sometimes the case that although one reported proof length is less than another,
the former proof will have steps with many more rewrites. This throws off the measure of
simplicity. (This is an issue only for \textsc{Prover9}, not other ATP systems.)
Some would argue that ideally, a simple proof should have no rewrites, with every step being
a primary inference \cite{WosPieper}.

\section{Final Remarks}

In this paper I have described a technique that can be used to simplify proofs generated
by automated theorem provers. This still begs important open questions already discussed earlier: what are
some reasonable measures of simplicity of a proof? Within the context of a particular theory, how can two
proofs be compared to decide which is the simpler? The techniques I have described have focused on one
particular measurement: the length of a proof as determined by the number of inferences. Within that context,
the simplification strategies described herein are certainly effective, but much more work is needed to
address the broader questions.


\begin{thebibliography}{99}

\bibitem{CSHPM} Canadian Society for the History and Philosophy of Mathematics,
    Program of the 2000 Annual Meeting at McMaster University, Ontario, Canada.
    \url{http://www.cshpm.org/archives/programs/prog2000.pdf}

\bibitem{Hillenbrand} Hillenbrand, Thomas,
    \textit{Waldmeister},
    Current version: February 2018.
    \url{https://www.mpi-inf.mpg.de/departments/automation-of-logic/software/waldmeister/}

\bibitem{KKP} Kinyon, Michael K., Kunen, Kenneth and Phillips, J. D.,
    Every diassociative A-loop is Moufang.
    \textit{Proc. Amer. Math. Soc.} \textbf{130} (2002), 619--624.

\bibitem{OTTER} McCune, William W.,
    \textit{OTTER (Organized Techniques for Theorem-proving and Effective Research)},
    Last supported version: 3.3f (August 2004).
    \url{http://www.cs.unm.edu/~mccune/otter/}

\bibitem{McCune} McCune, William W.,
    \textit{Prover9 and Mace4},
    Current version: 2009-11A.
    \url{http://www.cs.unm.edu/~mccune/prover9/}

\bibitem{Schulz} Schulz, Stefan,
    \textit{The E Theorem Prover},
    Current version: 2.0 Turzum (2017-07-04).
    \url{http://wwwlehre.dhbw-stuttgart.de/~sschulz/E/E.html}

\bibitem{Thiele1} Thiele, R\"{u}diger,
     Hilbert and his twenty-four problems,
     in \textit{Mathematics at the Dawn of a Millennium},
     Proceedings of the Canadian Society for History and Philosophy of Mathematics,
     Hamilton, Ont., 2000, vol. 13, M. Kinyon, ed., pp. 1--22.

\bibitem{Thiele2} Thiele, R\"{u}diger,
    Hilbert's twenty-fourth problem.
    \textit{Amer. Math. Monthly} \textbf{110} (2003), 1-–24.

\bibitem{ThieleWos} Thiele, R\"{u}diger and Wos, Larry,
    Hilbert's twenty-fourth problem.
    \textit{J. Automat. Reason.} \textbf{29} (2002), 67--89.


\bibitem{Veroff1} Veroff, Robert,
    Finding Shortest Proofs: An Application of Linked Inference Rules,
    \textit{J. Automat. Reason.} \textbf{27} (2001), 123--139.

\bibitem{Veroff2} Veroff, Robert,
    Solving Open Questions and Other Challenge Problems Using Proof Sketches,
    \textit{J. Automat. Reason.} \textbf{27} (2001), 157--174.

\bibitem{Voronkov} Voronkov, Andrei,
    \textit{Vampire},
    Current version: 3.0 (2018).
    \url{http://www.vprover.org/}

\bibitem{Wos} Wos, Larry,
    Automating the search for elegant proofs,
    \textit{J. Automated Reasoning} \textbf{21} (1998), 135--175.

\bibitem{WosPieper} Wos, Larry and Pieper, Gail W.,
    \textit{Automated reasoning and the discovery of missing and elegant proofs.}
    Rinton Press, Incorporated, Paramus, NJ, 2003.

\bibitem{WosThiele} Wos, Larry and Thiele, Ruediger,
    Hilbert's new problem.
    \textit{Bull. Sect. Logic Univ. \L\'{o}d\'{z}} \textbf{30} (2001), 165--175.

\end{thebibliography}
\end{document}